\documentclass[twocolumn,superscriptaddress,showpacs,amsmath,amssymb,nofootinbib]{revtex4-1}

\usepackage{graphicx}
\usepackage{dcolumn}
\usepackage{bm}
\usepackage{float}

\begin{document}

\title{Femtosecond Mega-electron-volt Electron Energy-Loss Spectroscopy}

\author{R. K. Li}
\email{lrk@slac.stanford.edu}
\author{X. J. Wang}
\affiliation{SLAC National Accelerator Laboratory, 2575 Sand Hill Road, Menlo Park, California, 94025, USA}

\date{\today}

\begin{abstract}
Pump-probe electron energy-loss spectroscopy (EELS) with femtosecond temporal resolution will be a transformative research tool for studying non-equilibrium chemistry and electronic dynamics of matter. In this paper, we propose a new concept of femtosecond EELS utilizing mega-electron-volt electron beams from a radio-frequency (rf) photocathode source. The high acceleration gradient and high beam energy of the rf gun are critical to the generation of 10-femtosecond electron beams, which enables improvement of the temporal resolution by more than one order of magnitude beyond the state of the art. The major innovation in our proposal - the `reference-beam technique', relaxes the energy stability requirement on the rf power source by roughly two orders of magnitude. Requirements on the electron beam quality, photocathode, spectrometer and detector are also discussed. Supported by particle-tracking simulations, we demonstrate the feasibility of achieving sub-electron-volt energy resolution and $\sim$10-femtosecond temporal resolution with existing or near-future hardware performances. 
\end{abstract}


\maketitle

\section{Introduction}

Electron energy-loss spectroscopy (EELS) analyzes the energy distribution of initially monoenergetic electrons after they have interacted with a specimen~\cite{eels, tem}. The change in kinetic energy of electrons carries rich information of the chemistry and electronic structures of the specimen atoms, which reveals the details of the bonding/valence states, the nearest-neighbor structures, the dielectric response, and the band gap, etc. EELS measurement, combined with diffraction in reciprocal space and imaging in real space using modern electron microscopes, provide a multi-dimensional panorama of material properties. 

An exciting new development of modern science focuses on the dynamics of material properties in non-equilibrium, such as heating, phase transitions, and chemical reactions, besides those in steady states~\cite{zewailxrayued, millerxrayued, fivechallenges, fivechallengesnew}. New X-ray~\cite{emma, ishikawa, lcls5years} and electron~\cite{king05, reed09, zewail10, germanued, browning12, flannigan12, musumeci:icfa,fesw, baumreview14} instruments and techniques with ever-improving temporal resolution, as well as spatial and energy resolutions are being developed to visualize these processes, aiming at fully understanding the connections between structures, dynamics, and functionality, and ultimately, controlling energy and matter.

Time-resolved EELS measurements have recently been carried out in ultrafast electron microscopes (UEM), which showcased their unique capabilities of mapping for example ultrafast electronic dynamics in solids and coherent quantum manipulation of free electrons in optical near-field~\cite{carboneeels09, carbonescience09, piazza14, flannigan14, caltecheels15, ropers15}. The technique, which is complementary to spectroscopy measurement using X-ray free-electron lasers (FEL), is well suited for very thin samples due to the much stronger interaction of electrons with material. Also, in principle, the probe size of electron beams can be focused by electromagnetic lenses to nanometer or smaller to provide detailed mapping of materials on atomic scales. 

Existing UEM instruments are based on modifying commercial transmission electron microscopes to operate with pulsed photoemitted electron beams, instead of the conventional continuous-wave thermionic or field-emission. It's highly challenging to reach desired energy and temporal resolutions in time-resolved EELS measurement, i.e. to minimize the energy spread and bunch length of electron beams simutaneously, which can both be severely degraded by electron-electron ($e$-$e$) interactions. The solution is to operate these instruments with extremely low charge density - on average a single electron per pulse, to eliminate the effects of $e$-$e$ interactions. The typical energy resolution of this operation mode is 1-2 eV~\cite{carboneeels09,carbonescience09,piazza14,caltecheels15,ropers15}. The temporal resolution, however, is still limited to several hundred femtoseconds, which are dominated by the pulse duration of the photoelectron beams. The photoelectrons are generated with a few tenths of eV initial energy spread, which translates into several hundred femtoseconds pulse duration due to vacuum dispersion~\cite{flannigan14,ropers15}. While at least one-order-of-magnitude shorter bunch length is desired to capture fast dynamics of electronic structures. 

To tackle the challenge associated with vacuum dispersion, an electron source with a significantly higher acceleration gradient and higher output energy would be necessary. Radio-frequency (rf) photocathode guns, featuring 10s to 100 MV/m gradient and several MeV beam energy, would be the ideal choice. Recently, these sources have been optimized for ultrafast electron diffraction~\cite{xj03,xj06,slacued06,ucla08,lrk09,osaka11,bnl13,jiaoda14,regae, astaued, daresbury, kaeri, hires} and imaging~\cite{lrkuem14,xduem14,ucla10fs} with transformative impacts by delivering unprecedented temporal and reciprocal-/real-space resolutions. Unfortunately, the energy stability of the electron beams from rf guns, determined by the stability of the driving rf power sources, is currently at best at $1\times10^{-5}$ level, i.e. 50 eV for 5 MeV beams, which is far from adequate for spectroscopy applications. Thus unless some technical breakthrough can improve the rf stability by at least two orders of magnitude, it was regarded impractical to consider photocathode rf guns for fs EELS. 

Here we propose a `reference beam technique' that significantly relaxes the demanding requirement on the rf stability, and demonstrate the feasibility of fs MeV EELS based on rf photocathode electron sources. We will present a complete conceptual design of a fs MeV EELS system. Detailed simulation results show that one can achieve sub-eV energy resolution even with ~50 eV beam energy fluctuation. Meanwhile, the temporal resolution is improved to the 10-fs level, which is more than one order of magnitude beyond the state of the art. One may also take advantage of the MeV beam energy to study thicker samples, due to the reduced inelastic cross section compared to commonly used 200-300 keV electrons. Or, for the same sample thickness there will be reduced multiple scattering with MeV electrons, which significantly simplifies the interpretation of the EELS spectrum.  

In this paper, we will first introduce the concept of the `reference beam technique' and the overall system design in Section~\ref{sec2}. The temporal resolution, which includes contributions from the electron beam pulse duration and pump-to-probe time-of-arrival jitter, will be evaluated in Section~\ref{sec3}. Impacts of the gun rf field on the energy resolution will be presented in Section~\ref{sec4}. We will discuss the requirements on the spectrometer resolution and transverse beam qualities in Section~\ref{sec5}. The effects of $e-e$ interaction on the energy resolution will be discussed in Section~\ref{sec6}.

\section{`Reference beam' concept for fs MeV EELS}
\label{sec2}

\begin{figure*}[ht]
\includegraphics[width=0.7\textwidth]{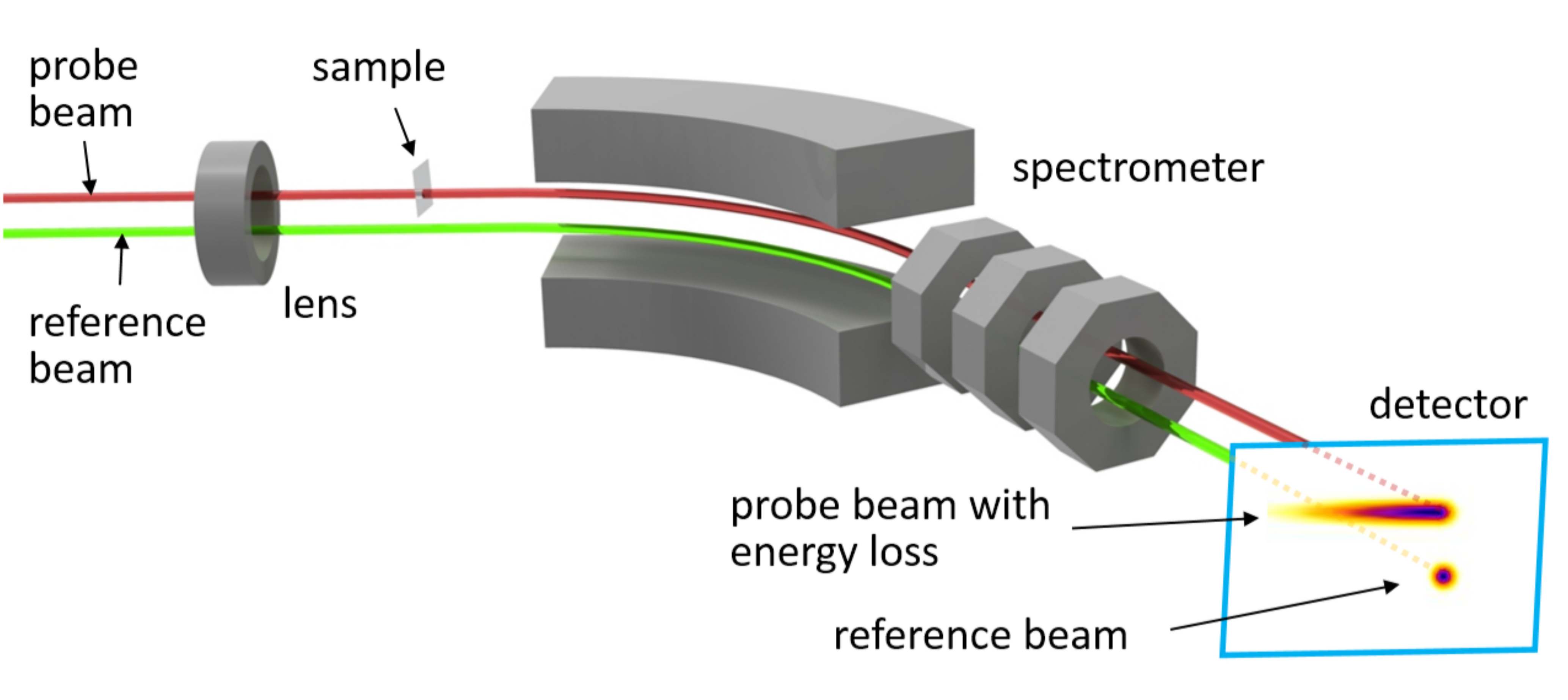}
\caption{(Color online) Concept of the `reference beam' technique for fs MeV EELS based on rf photocathode guns. }
\label{fig:EELSconcept}
\end{figure*}

We illustrate in Fig.~\ref{fig:EELSconcept} the concept of `reference beam technique' for fs MeV EELS based on an rf photocathode electron source. Two electron beams, which are called the `probe beam' and `reference beam' respectively, are generated from the photocathode with both transverse ($\sim$100 $\mu$m) and longitudinal (time, $\sim$1 picosecond) separations. The energies of both electron beams will fluctuate at 50-eV level ($1\times10^{-5}$ of 5 MeV) due to the stability of the rf power source. However, the {\it difference} between their energies and the energy spread of each individual beam can all be controlled at sub-0.1-eV level. Detailed analysis on the contribution from the gun rf field and $e$-$e$ interactions will be presented in Section~\ref{sec4} and \ref{sec6}, respectively. 

At the sample location the two beams are also transversely (vertically) separated, and only the `probe beam' will interact with the sample. The longitudinal separation is necessary to minimize $e$-$e$ interactions at transverse focus along the beamline. A high resolution spectrometer will measure the energy of both the scattered `probe beam' and the unperturbed `reference beam'. The energy difference between the two beams consists of two parts: (1) the energy loss due to the sample, and (2) the original energy difference when the sample is not present. By recording the probe and reference beam pair on shot-by-shot basis and comparing the energy difference, one can construct a complete energy-loss spectrum due to the sample. Note that the original energy difference contributes as a fixed offset of the zero-loss peak and doesn't distort of the energy axis of the spectrum. 

The energy resolution of the `reference beam technique' is 
\begin{equation}
\Delta E^2=(E_\textrm{probe}-E_\textrm{ref})^2+\delta E_\textrm{probe}^2+(\sigma/D)^2,
\end{equation}
where $E_\textrm{probe/ref}$ is the average energy of the probe/reference beam, $\delta E_\textrm{probe}$ is the energy spread of the probe beam. Note that the energy spread of the reference beam doesn't contribute here since only its average energy is relevant. $\sigma/D$ is the instrumentation energy resolution of the spectrometer, where $\sigma$ is the beam spot size on the spectrometer detector, and $D$ is the spectrometer dispersion. To ensure that $\sigma/D$ is also well below 0.1 eV, with a practical spectrometer design $\sigma$ should be less than a few tens of nm on the sample. Combined with the requirement on the beam divergence, which should be much smaller than the typical Bragg angle of ~1 mrad for 5 MeV electrons, the normalized beam emittance should be sub-nm-rad.  

The success of fs MeV EELS will rely on the generation and preservation of sub-eV energy spread, ~10 fs bunch length, and sub-nm emittance electron beams from the source, through the sample, till the spectrometer. Such a precisely shaped and miniature phase space volume can accommodate only a single electron per pulse, as will be shown in Sec.~\ref{sec6}. A high repetition-rate electron source is a natural choice to build up high signal-to-noise ratio within a reasonable data acquisition time. In the circumstances where the pump-probe repetition rate is limited to kHz due to laser-induced sample heating, pulsed rf gun is also an option with $>$100 MV/m acceleration gradient to deliver shortest possible electron beam pulse durations. The simulation results presented in following Sections are based on the design of a 200 MHz quarter-wave resonator type superconducting rf (SRF) gun.

\begin{figure}[htb]
\includegraphics[width=0.45\textwidth]{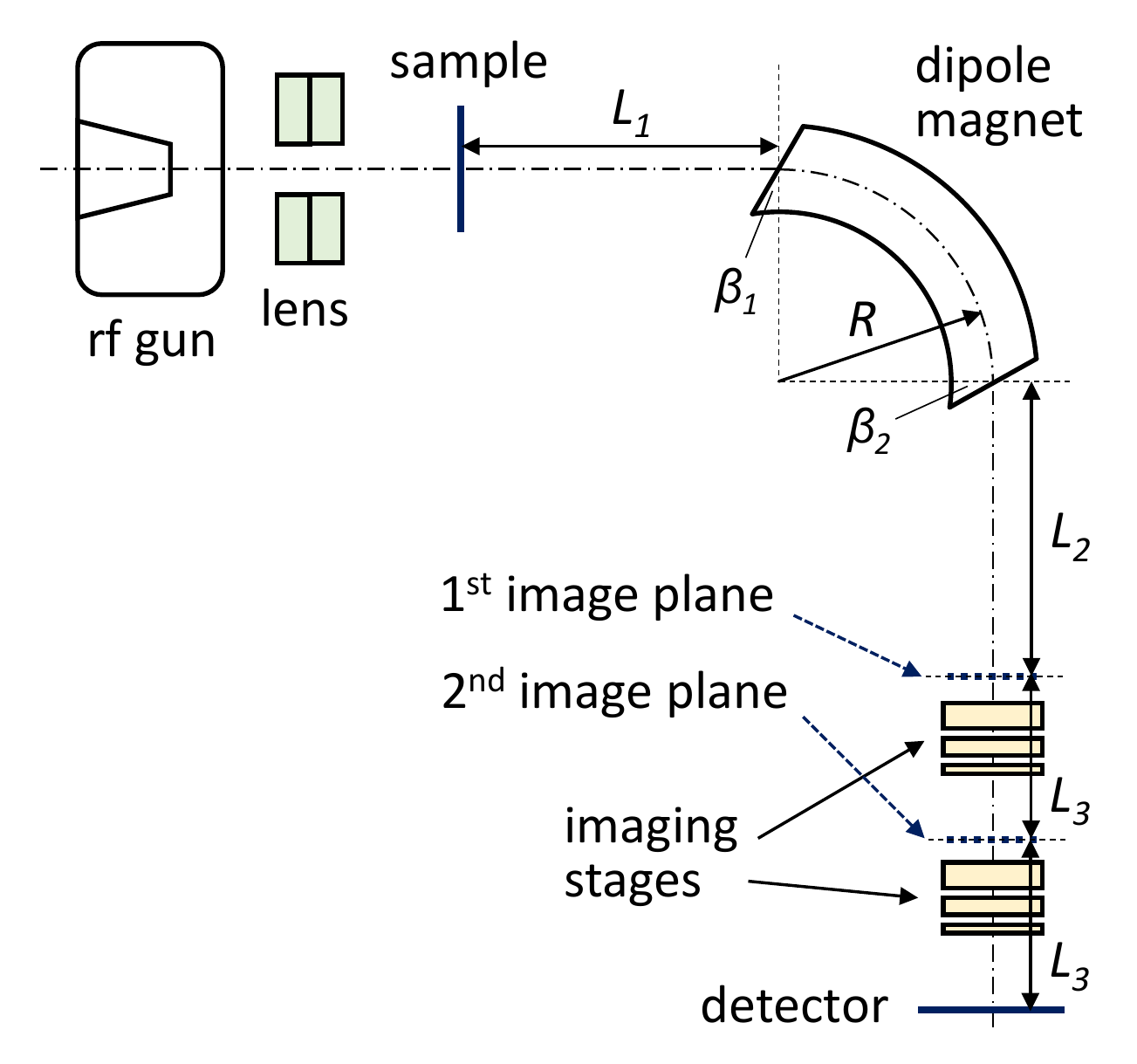}
\caption{(Color online) Schematic of the fs MeV EELS system. Components include an rf gun, a condenser lens, the sample, and a spectrometer. The spectrometer consists of a dipole magnet and two stages of magnifying imaging lenses.}
\label{fig:techlayout}
\end{figure}

A more technical schematic of the system is shown in Fig.~\ref{fig:techlayout}, which includes an rf gun, a condenser lens, and a high resolution spectrometer, etc. The design and consideration of each key component, as well as the control and evolution of the electron beams, will be discussed in following Sections.

\section{Temporal resolution: bunch length and Time-of-arrival jitter}
\label{sec3}

In a pump-probe EELS measurement, a pump laser pulse first illuminates the sample and prepares the system to an excited state. After a controlled time delay a probe electron beam interacts with the samples and captures the transient electronic property. By repeating the pump-probe events at various time delay, one can reconstruct the full evolution of the dynamic process. The temporal resolution of the measurement is 
\begin{equation}
\tau=(\tau_e^2+\tau_\textrm{pump}^2+\tau_\textrm{TOA}^2+\tau_\textrm{VM}^2)^{1/2},
\end{equation}
where $\tau_e$ and $\tau_\textrm{pump}$ are the pulse durations of the probe electron and pump laser pulses, respectively, $\tau_\textrm{TOA}$ is the time-of-arrival (TOA) jitter between the pump and probe pulses at the sample, and $\tau_\textrm{VM}$ is the velocity mismatch term~\cite{vm93}. $\tau_\textrm{VM}$ is negligible for $\mu$m and thinner samples. 

\begin{figure}[htb]
\includegraphics[width=0.45\textwidth]{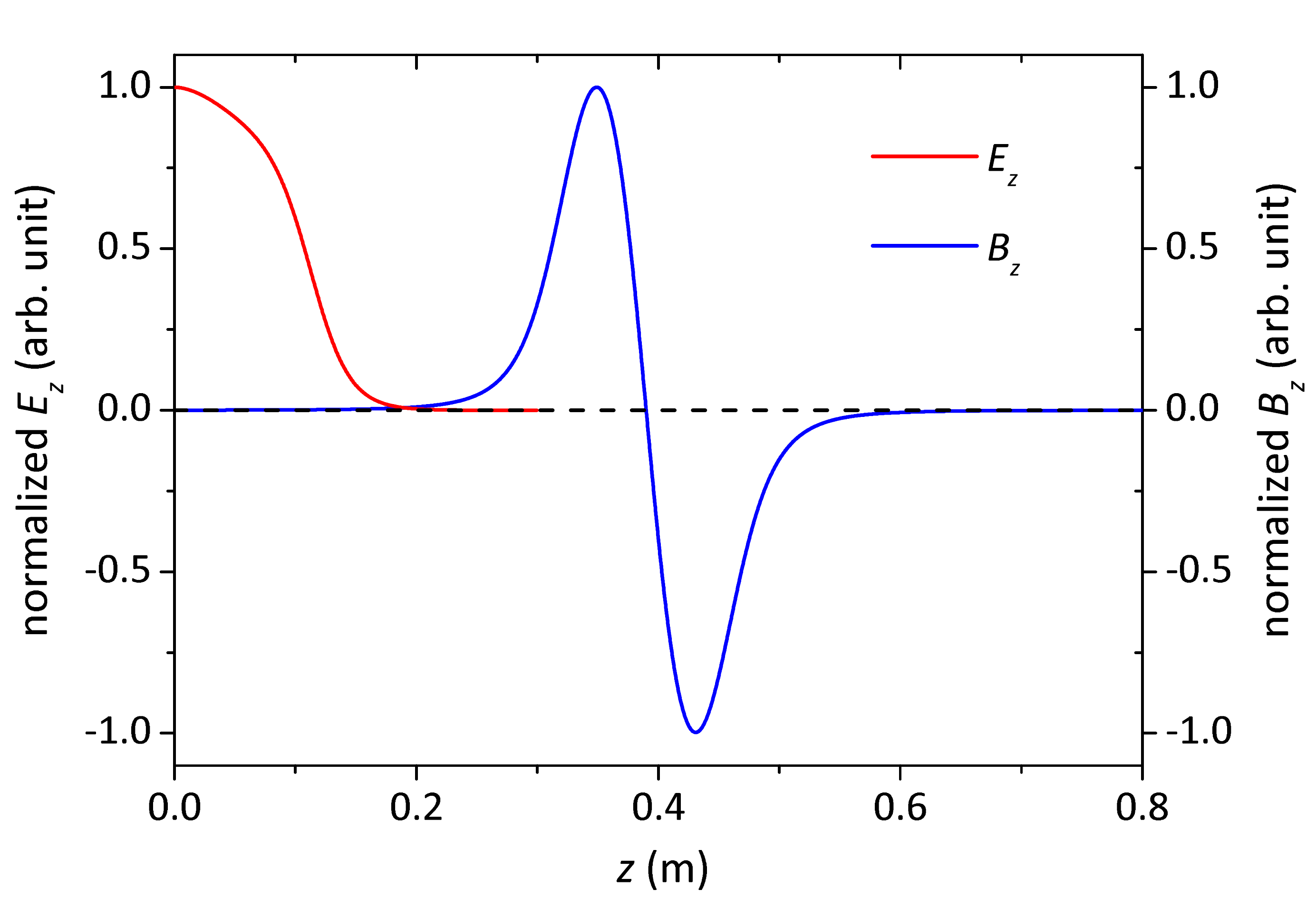}
\caption{(Color online) On-axis longitudinal electric field $E_z$ of the rf gun and on-axis longitudinal magnetic field $B_z$ of the solenoid condenser lens.}
\label{fig:fieldmap}
\end{figure}

$\tau_e$ is determined by several factors, including the longitudinal dynamics in the rf field, initial energy spread induced vacuum dispersion, as well as the initial pulse duration from photoemission. The longitudinal dynamics in an rf gun depends on the field strength and the launch phase when the photoelectrons are generated. The on-axis longitudinal electric field $E_z$ of the rf gun is shown in Fig.~\ref{fig:fieldmap}. Here we have to choose a launch phase for close to maximum output energy hence minimal rf-induced energy spread, to eventually reach high EELS energy resolution. The reason will be discussed more quantitatively in Section.~\ref{sec4}. With this launch phase, there is no effective rf compression~\cite{xj96} and $\tau_e$ is dominated by the initial pulse duration and vacuum dispersion. 

The initial energy distribution of photoelectrons consists of several parts, including the thermal spread due to the finite electronic temperature of the cathode,  the excitation bandwidth of the cathode driving laser, and the mismatch between the photon energy and the effective work function~\cite{Jensen07, Ang08, dowell09, psi10, baumsinglee10, luca15}. One can shift the central wavelength to minimize the last mismatch term. The effect of laser induced cathode heating can be controlled at negligibly small level in our extremely low bunch charge regime~\cite{Jared16}. In an SRF gun the cathode temperature is well below room temperature (25 meV), and we assume the thermal spread is 5 meV rms in both transverse and longitudinal directions. The minimal value of the laser excitation bandwidth is given by the Fourier transform limit. For an Gaussian temporal profile pulse, the FWHM excitation bandwidth is $\delta E_\textrm{laser} \textrm{[meV]}=1822/\tau_\textrm{laser}\textrm{[fs]}$. 

\begin{figure}[htb]
\includegraphics[width=0.45\textwidth]{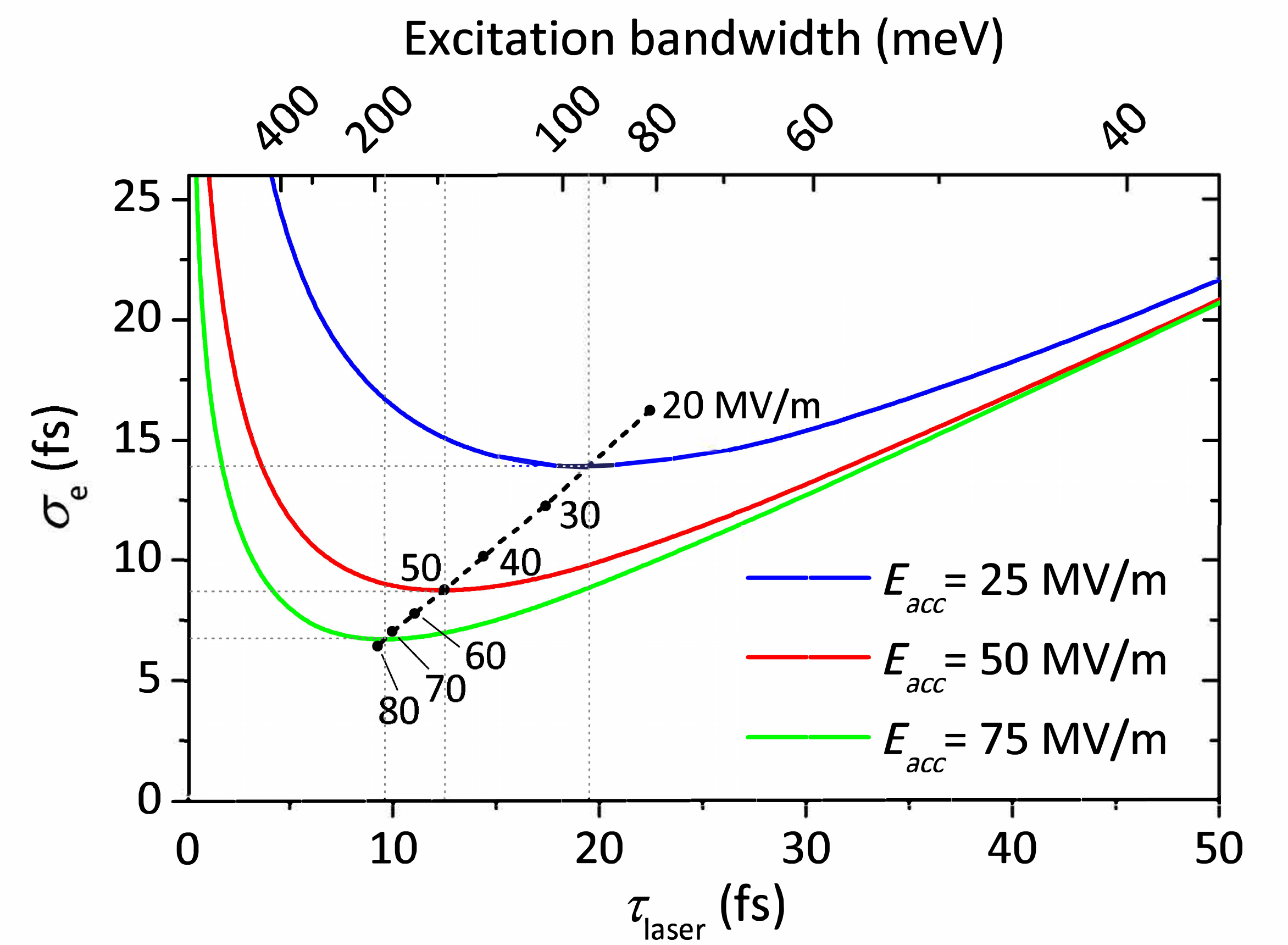}
\caption{(Color online) The electron beam pulse duration at the sample location $\sigma_e$ can be optimized by tuning the pulse duration of the cathode-driving laser $\tau_\textrm{laser}$. Here we assume the initial pulse duration of the electron beam is equal to $\tau_\textrm{laser}$ with a prompt photocathode, and the initial energy spread of the electron beam includes contribution from the excitation bandwidth of a transform-limited laser. Higher rf gun gradient enables shorter minimal $\sigma_e$. The black dots indicate the minimal $\sigma_e$ for gun gradient from 20 MV/m to 80 MV/m.}
\label{fig:pulseduration}
\end{figure}

In order to generate short $\sigma_e$, since there is no effective rf compression and vacuum dispersion will only lengthen the beam, it is important to start with short initial pulse length, and hence a photocathode with prompt response is highly desired. We choose metallic cathodes with a few fs response time and assume the initial pulse duration of the electron beam approximately equals to that of the driving laser. On the other hand, shorter driving laser is associated with a larger, transform-limited excitation bandwidth, and leads to excessive electron beam lengthening. In Fig.~\ref{fig:pulseduration}, we show that $\sigma_e$ can be optimized by adjusting $\tau_\textrm{laser}$ to balance the two competing effects. Higher rf field gradient and beam energy can more effectively suppress vacuum dispersion and generate shorter $\sigma_e$. We performed the simulation using the General Particle Tracer code~\cite{gpt}. In the rest of the paper we will focus on the 50 MV/m gradient case, where $\sigma_e$ is below 10 fs rms.

\begin{figure}[htb]
\includegraphics[width=0.45\textwidth]{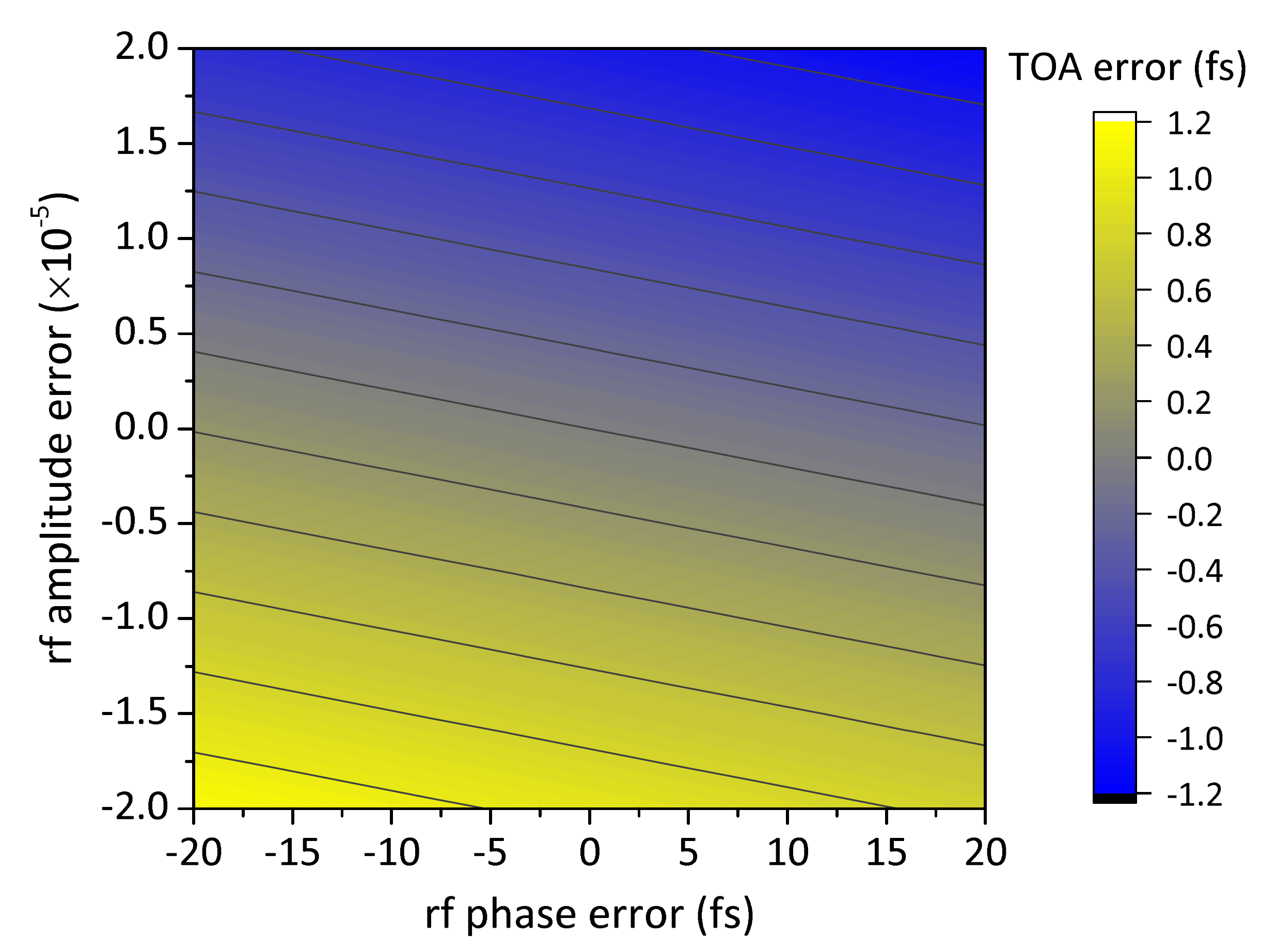}
\caption{(Color online) TOA error between the pump laser and probe electron beams at the sample due to the phase and amplitude fluctuations of the gun rf field.}
\label{fig5}
\end{figure}

The other important contributing term $\tau_\textrm{TOA}$ is determined by the phase and amplitude jitters of the rf field. Here we assume the cathode drive laser and the sample pump laser are split from a common laser pulse and thus essentially jitter-free. TOA error can be evaluated in a straightforward way by adding small errors in the rf amplitude and launch phase to the nominal settings. We assume that the rf amplitude and phase errors are $1\times10^{-5}$ rms and 10 fs rms, respectively, which are typical for a state-of-the-art CW source. The TOA error stays below ~1 fs within the range of $-\sigma$ to $+\sigma$ for both phase and amplitude errors, as shown in Fig.~\ref{fig5}. Note that since at this launch phase there is minimal rf compression effect, i.e. the TOA is not sensitive to the rf phase jitter, the TOA depends more strongly on the rf amplitude fluctuation. 

In this section, we have quantitatively evaluated the main contributing terms to the temporal resolution, and demonstrated that $\tau_e$ and $\tau_\textrm{TOA}$ can be controlled at 10 fs and 1 fs level, respectively. The pulse duration of the pump laser can be readily maintained at 10 fs level, and the velocity mismatch term is negligible for solid state samples, which are $\mu$m or thinner. Hence we conclude that the overall temporal resolution of fs MeV EELS is at the 10-fs level.  

\section{rf-field contribution to the energy resolution}
\label{sec4}

For an rf gun, due to the spatial and temporal variation of the rf electromagnetic field, the output energy of each photoelectron depends its particular trajectory through the field. The trajectory is determined by its initial position and angle from the photocathode and the launch phase, i.e. the rf phase at the instance of photoemission. The accelerating field in the rf gun is axially symmetric around the beam axis ($z$ axis). The longitudinal and transverse components are 
\begin{eqnarray}
E_z&=&E_\textrm{acc}e(z,r)\sin(\omega_0t+\phi)\\
E_r&=&-\frac{r}{2}\frac{\partial E_z}{\partial z}
\label{rffield}
\end{eqnarray}
where $E_\textrm{acc}$ is gun gradient, $e(z,r)$ is the normalized field profile, $\omega_0$ is the resonant frequency, and $\phi$ is the launch phase. It is straightforward to calculate the average energy and energy spread of an electron beam for given initial spot position, spot size, divergence, and pulse duration. The main parameters are summarized in Table.~\ref{tab1}. 

\begin{table}[h]
\caption{\label{tab1} Main machine and beam parameters for fs MeV EELS.}
\begin{ruledtabular}
\begin{tabular}{lc}
Parameters & Values \\
\hline
gun gradient & 50 MV/m \\
gun frequency & 200 MHz \\
launch phase for max. output energy & 73.83 degree \\
max. output energy & 5.12 MeV\\
solenoid strength, $B_0$ & 0.40 T \\
beam charge & $<$1$e$/pulse \\
initial spot size, rms (uniform) & 50 nm \\
intrinsic emittance & 0.23 $\mu$m/mm\\
initial pulse duration, FWHM & 18.2 fs \\
transverse offset $\Delta y_\textrm{probe/ref}$ & $\pm$50 $\mu$m \\
temporal offset $\Delta t_\textrm{probe/ref}$ & $\pm$0.5 ps \\
\multicolumn{2}{l}{\bf At the sample ($z=70$ cm)} \\
horizontal beam centroid $x_\textrm{0,probe/ref}$ & 0/0 $\mu$m \\
horizontal beam size, rms $\sigma_{x,\textrm{probe/ref}}^{rms}$ & 13.8/13.8 nm \\
horizontal beam size, FW50 $\sigma_{x,\textrm{probe/ref}}^\textrm{FW50}$ & 24.0/24.0 nm \\
vertical beam centroid $y_\textrm{0,probe/ref}$ & -13.8/13.8 $\mu$m \\
vertical beam size, rms $\sigma_{y,\textrm{probe/ref}}^{rms}$ & 13.4/13.4 nm \\
vertical beam size, FW50 $\sigma_{y,\textrm{probe/ref}}^\textrm{FW50}$ &  22.7/22.6 nm \\
beam divergence, rms $\sigma_{x'/y',\textrm{probe/ref}}$ & 76 $\mu$rad \\
normalized emittance, rms $\epsilon_{{x/y},\textrm{probe/ref}}$ & 11.5 pm-rad \\
temporal separation $t_{0,\textrm{probe}}-t_{0,\textrm{ref}}$ & 1.0 ps \\
bunch length, rms $\sigma_{t,\textrm{probe/ref}}$ & 9.2/9.4 fs \\
beam energy $E_\textrm{probe/ref}$ & 5.12 MeV \\
energy difference $E_\textrm{probe}-E_\textrm{ref}$ & 0.001 eV \\
energy spread, FW50 $\delta E_\textrm{probe/ref}$ & 0.13/0.15 eV \\
\end{tabular}
\end{ruledtabular}
\end{table}

\begin{figure}[htb]
\includegraphics[width=0.45\textwidth]{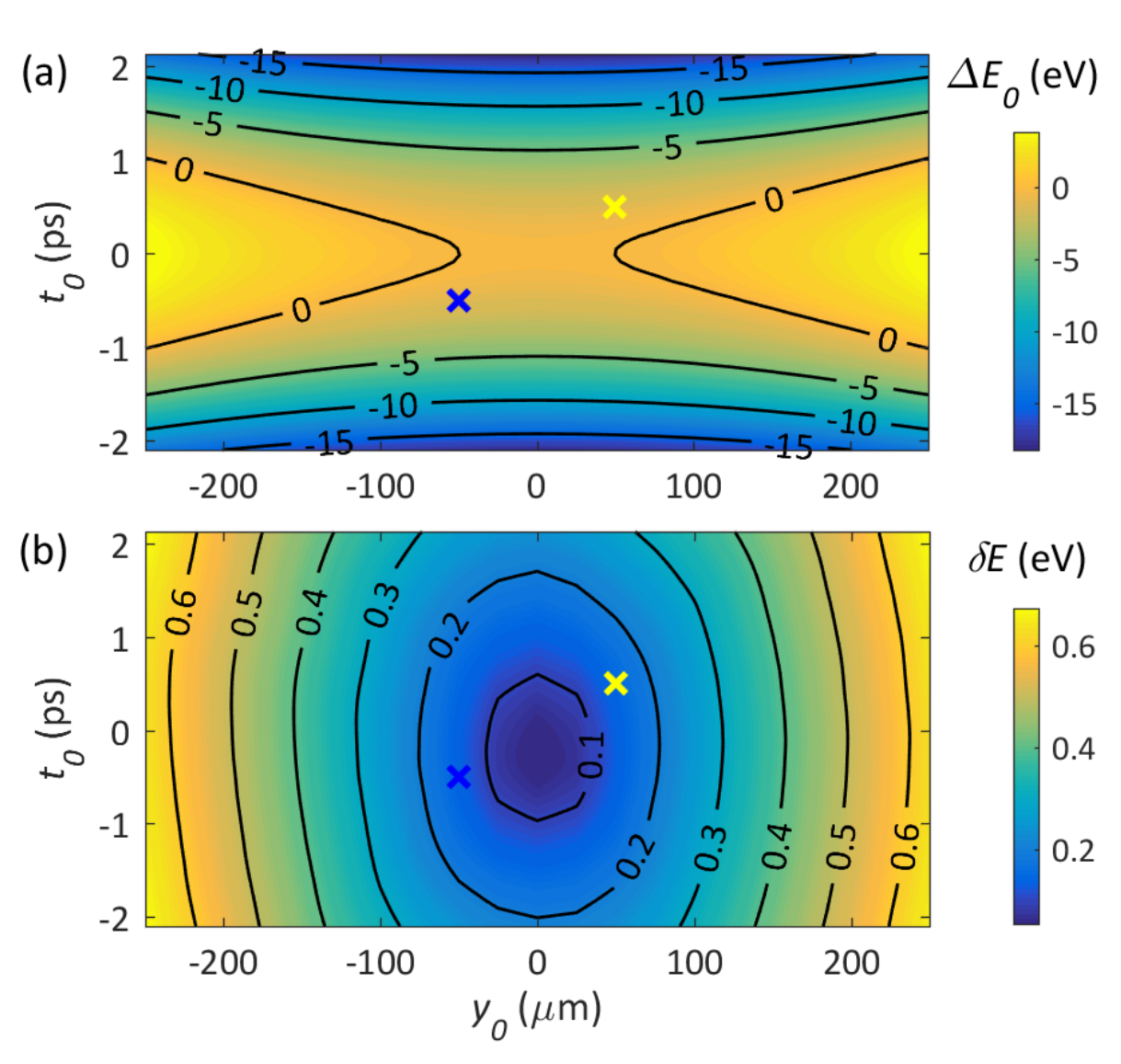}
\caption{(Color online) Dependence of (a) the average energy and (b) FW50 energy spread of an electron beam on its initial vertical offset from the gun center $\Delta y$ and temporal offset $\Delta t$ from the launch phase for maximum output energy. The probe and reference beams are indicated by the yellow and blue cross, respectively.}
\label{fig:edephaseampscan}
\end{figure}

In Fig.~\ref{fig:edephaseampscan}, we show the particle tracking results for the average energy and full-width 50\% (FW50) energy spread of an electron beam for various initial transverse offset $\Delta y$ and temporal offset $\Delta t$, with other parameters as specified in Table.~\ref{tab1}. Here $\Delta y$ is defined relative to the gun center, and $\Delta t$ is with respect to the launch phase for maximum output energy. $\Delta y$ and $\Delta t$ should be large enough such that the $e$-$e$ interactions between the probe and reference beams are negligible. We will discuss in detail in Section.~\ref{sec6} the effects of $e$-$e$ interactions, and $\Delta y=\pm50~\mu$m and $\Delta t=\pm0.5$ ps are found adequate. The probe beam and reference beam are located at $(\Delta y, \Delta t)$ and $(-\Delta y, -\Delta t)$, respectively. With these separations, the difference between the average energy of the probe and reference beams is less than 1 meV, and their energy spread are both controlled below 0.15 eV. 

\begin{figure}[htb]
\includegraphics[width=0.45\textwidth]{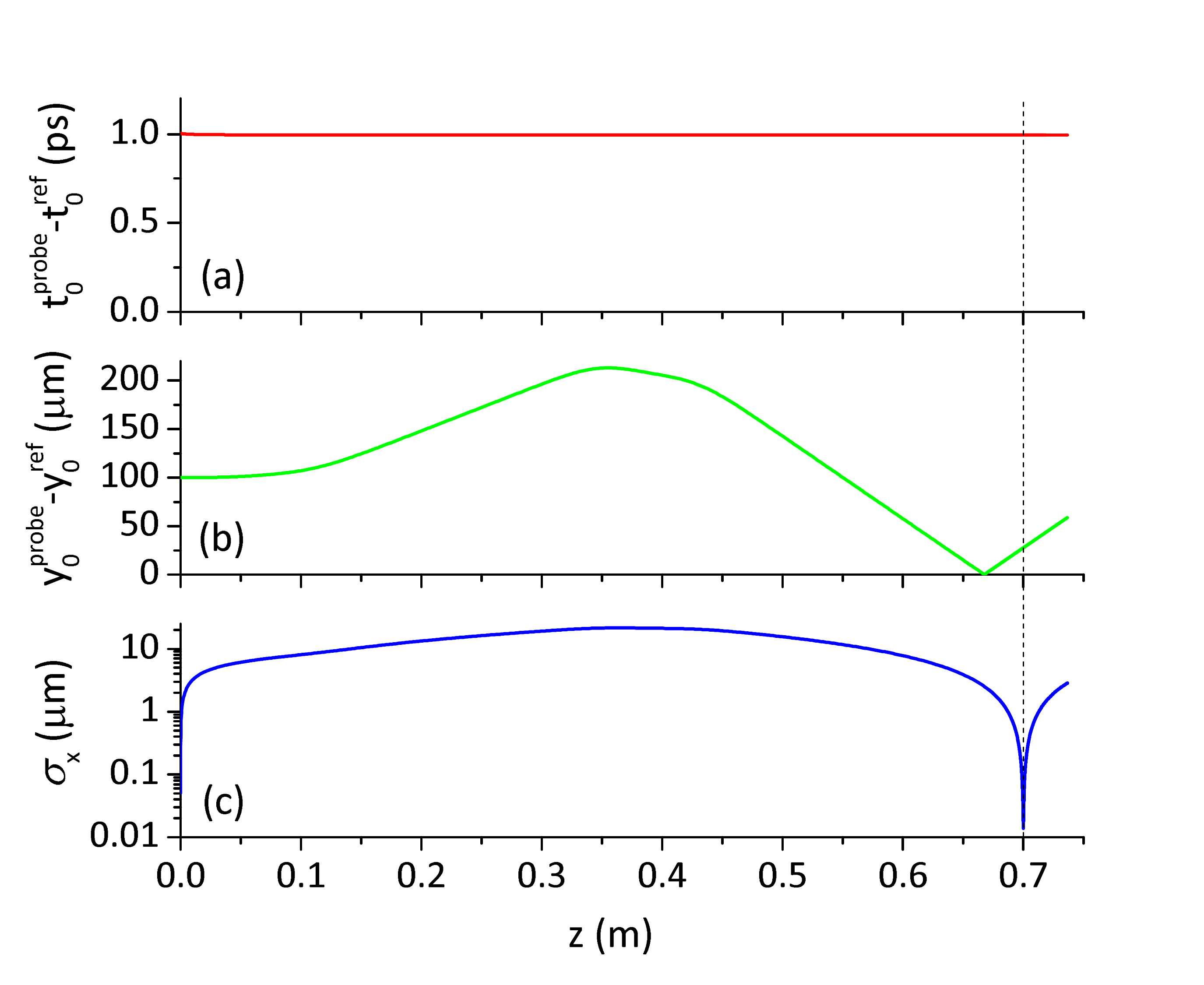}
\caption{(Color online) The (a) longitudinal (temporal) and (b) transverse separations between the probe and reference beams, as well as the (c) rms transverse beam size from the photocathode ($z=0$ m) to the sample location ($z=0.7$ m).}
\label{fig:y_t_diff}
\end{figure}

The initial offsets $\pm\Delta y$ and $\pm\Delta t$, together with the rf gun and solenoid configuration, control the longitudinal and transverse separations between the probe and reference beams along the beam line. The solenoid condenser lens, as shown in Fig.~\ref{fig:fieldmap}, consists of two identical coils with opposite polarities so that the integrated rotation of the electron beams is zero as the kinetic energy stays constant through the lens. In Fig.~\ref{fig:y_t_diff} we show the longitudinal and transverse separations, as well as the transverse beam size from the photocathode to the sample location ($z=0.7$ m). The temporal separation stays constant at 1 ps since the relative longitudinal particle motion is essentially frozen. The two beams are 27.6 $\mu$m apart in $y$ direction at the sample location, and only the probe beam will interact with the sample. The sample plane will be imaged to the spectrometer detector with a total magnification of $16\times16$ times in dispersion ($x$) direction and $4\times4$ times in $y$ direction. The design of the spectrometer will be discussed in the Section~\ref{sec5}.

\begin{figure}[htb]
\includegraphics[width=0.45\textwidth]{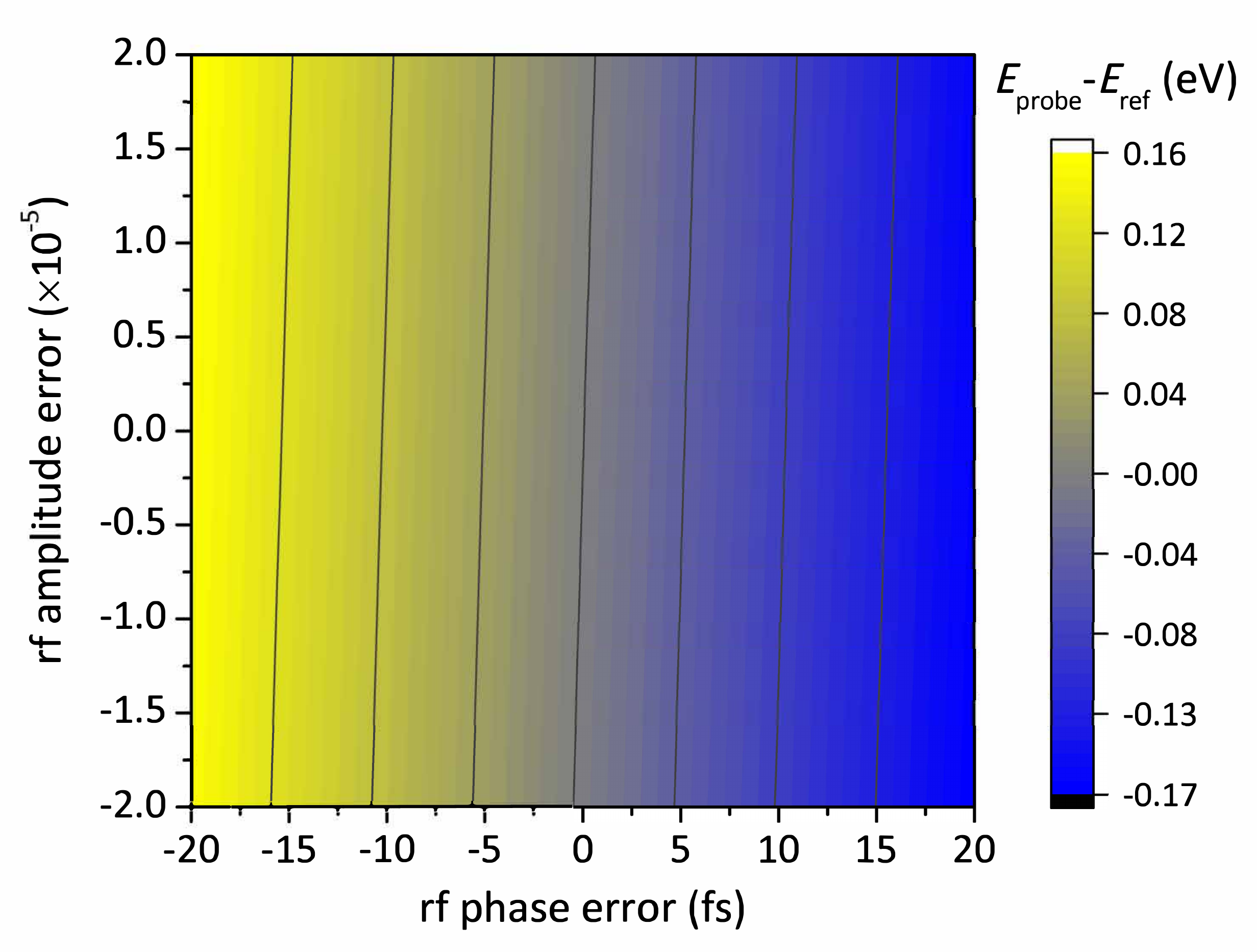}
\caption{(Color online) Dependence of the energy difference between of the probe and reference beams $E_\textrm{probe}-E_\textrm{ref}$ on the phase and amplitude errors of the gun rf field. The rf phase and amplitude fluctuations are 10 fs rms and $1\times10^{-5}$ rms, respectively.}
\label{fig:ediffscoff}
\end{figure}

Due to the timing error of the rf launch phase relative to the cathode driving laser $\Delta t_\phi$, the temporal offsets of the probe and reference beams become $\Delta t_\phi+\Delta t$ and $\Delta t_\phi-\Delta t$, respectively, which leads to difference between their average energies. The dependence of the energy difference $E_\textrm{probe}-E_\textrm{ref}$ on the rf phase error, as well as on the rf amplitude error, is shown in Fig.~\ref{fig:ediffscoff}. It is evident that the energy difference is insensitive to the rf amplitude fluctuation. This results quantitatively demonstrate the effectiveness of the `reference beam technique'. With assumed specifications of 10 fs rms rf phase error and $1\times10^{-5}$ rms amplitude error, the uncertainty of the energy difference is 80 meV rms, while the energy spread of each individual beam changes less than 1 meV. 

\section{spectrometer resolution, beam emittance requirement, and photocathode solution}
\label{sec5}

In order to precisely model the $e$-$e$ interaction between the probe and reference beam from the cathode to the detector, a complete beam optics design, including the spectrometer with imaging optics, is required to define the beam trajectory and envelop. A first-order spectrometer design was illustrated in Fig.~\ref{fig:techlayout}. The spectrometer consists of a double-focusing dipole magnet followed by two stages of magnifying imaging optics. The layout and main parameters of the spectrometer are summarized in Table.~\ref{tab2}. The dipole bends the beam trajectory with a radius of 0.5 m and an angle of $\pi/2$. With tilted pole-faces at both the entrance and exit, the beam is imaged in both $x$ and $y$ directions with a 1:1 magnification from the object plane (sample) to the first image plane. Since the beam spot size is only a few tens of nm on the first image plane, imaging optics is necessary to magnify the beam spot to match the point spread function of the detector. Here we choose two identical imaging stages. Each stage consists of a permanent magnet triplet~\cite{lrkuem14, pmqprl16} which magnifies 16.0 times in $x$ (the horizontal and dispersion) direction and 4.0 times in $y$ direction. 

\begin{table}[h]
\caption{\label{tab2} Main parameters of the spectrometer.}
\begin{ruledtabular}
\begin{tabular}{lc}
Parameters & Values \\
\hline
bending radius $R$ & 0.5 m \\
bending angle & $\pi/2$ \\
dipole strength, $B_0$ & 0.374 kG \\
pole-face tilt angle, $\beta_1$ and $\beta_2$ & $27.3^{\circ}$ \\
sample to dipole entrance, $L_1$ & 1.534 m \\
dipole exit to $1^\textrm{st}$ image plane, $L_2$ & 1.534 m \\
$1^\textrm{st}$ to $2^\textrm{nd}$ image plane, $L_3$ & 0.15 m \\
$2^\textrm{nd}$ image plane to detector, $L_3$ & 0.15 m \\
\end{tabular}
\end{ruledtabular}
\end{table}

\begin{table}[h]
\caption{\label{tab3} 
Parameters of the PMQs for a single triplet imaging stage. The object plane is at $z=0$ and the image plane at $z=0.15$ m.}
\begin{ruledtabular}
\begin{tabular}{ccccc}
Name & Thickness & Gradient & Position \\
\hline
$Q_1$ & 6 mm & 537.5 T/m & 8.51 mm \\
$Q_2$ & 4 mm & -537.5 T/m & 22.07 mm \\
$Q_3$ & 2 mm & 537.5 T/m & 28.92 mm \\
\end{tabular}
\end{ruledtabular}
\end{table}

\begin{figure}[htb]
\includegraphics[width=0.45\textwidth]{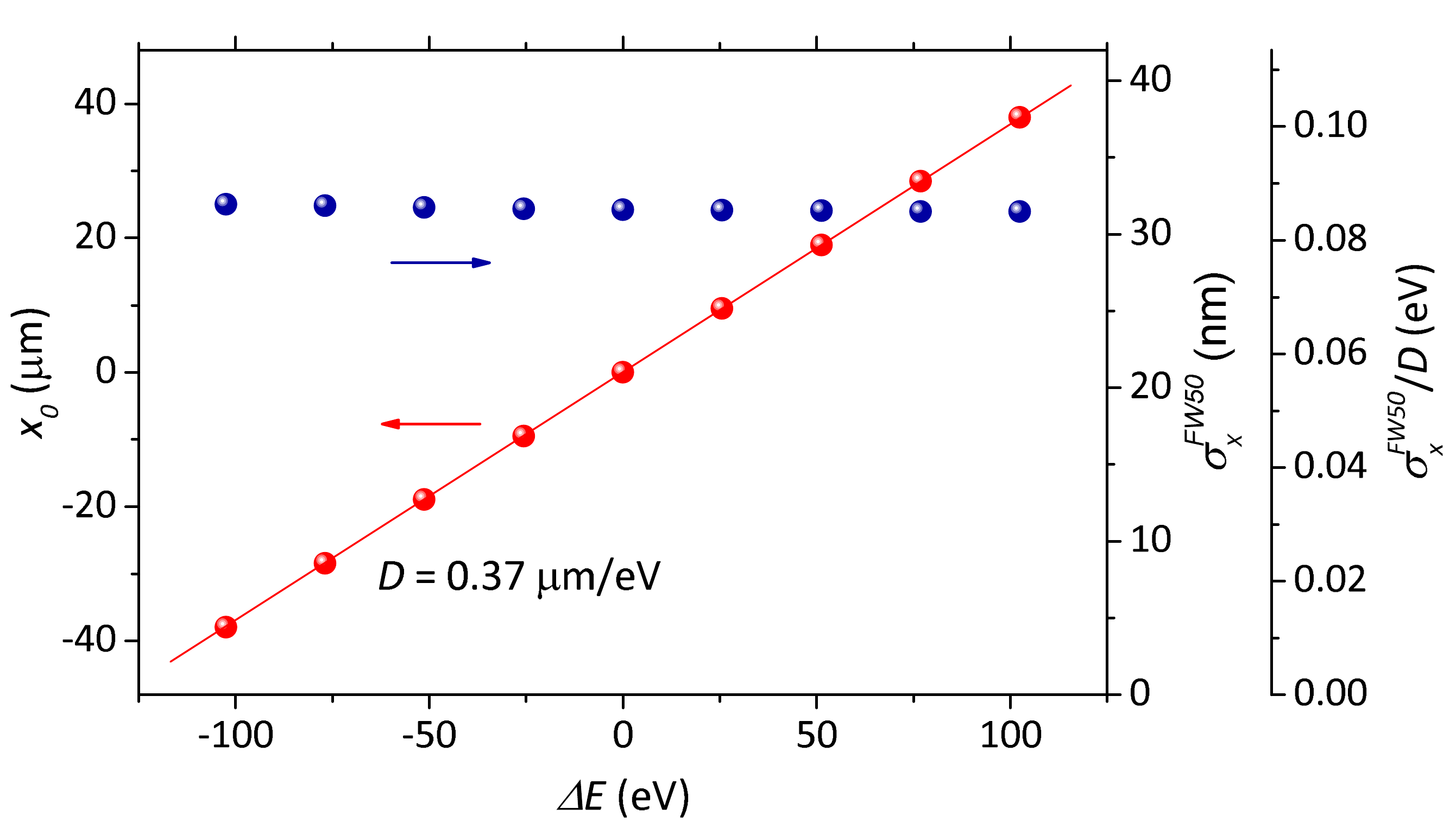}
\caption{(Color online) At the first image plan, the beam centroid in the horizontal direction $x_0$ for beam energies off the nominal value by $-2\times10^{-5}$ to $2\times10^{-5}$, or roughly -100 to 100 eV for 5 MeV beam energy. The beam size $\sigma_x^\textrm{FW50}$ stays approximately 32 nm, which corresponds to 85 meV.}
\label{fig:dispersion}
\end{figure}

The horizontal beam size on the first image plane $\sigma_x^{\textrm FW50}$ includes contributions from the dipole dispersion, the transverse beam size at the sample, and possible aberrations in imaging. Thus $\sigma_x^{\textrm FW50}/D$ is the upper bound of the beam energy spread, where $D$ is the dispersion at the first image plane. In Fig.~\ref{fig:dispersion}, we show the horizontal beam centroid $x_0$ and beam size $\sigma_x^{\textrm FW50}$ on the first image plane for energy variation of -100 to 100 eV ($-2\times10^{-5}$ to $-2\times10^{-5}$ of 5 MeV) around the nominal value. The dispersion, i.e. the slope of the centroid curve, is $D=0.37$ $\mu$m/eV. $\sigma_x^{\textrm FW50}$ is maintained around 32 nm, which corresponds to 85 meV. The electron beam parameters are listed in Table.~\ref{tab1}, except with $\Delta y=0$ and $\Delta t=0$, and the actual FW50 energy spread is $\delta E=62$ meV. The results has demonstrated that the spectrometer is capable of resolving $<$0.1 eV energy spread. 

The undesired components in the horizontal beam size for energy spread determination, including contributions from the beam spot size at the sample and possible aberration in imaging, are related to the emittance of the electron beam. To allow characterization of sub-0.1 eV energy spread with a dispersion of $D=0.37$ $\mu$m/eV, the horizontal beam size needs to be less than 37 nm on the first image plane, hence also at the object plane (sample location) with a 1:1 imaging. Meanwhile, the constrain we choose for the beam divergence at the sample $\sigma^\textrm{FW50}_{x'/y'}$ is that it should be one order of magnitude smaller than the typical Bragg angle ($\sim$1 mrad for 5 MeV electrons), thus the divergence should be $\sim$100 $\mu$rad or smaller. Combining the two aspects, the upper limit for the normalized FW50 beam emittance is 40 pm-rad, or roughly 25 pm-rad with the rms definition. 

It is a challenge but actually feasible to generate 25 pm-rad or lower emittance from a photocathode. The intrinsic emittance is estimated to be 0.23 mm-mrad per mm rms emission size, which is dominated by the laser excitation bandwidth. The contribution from the cathode temperature is negligible, since the cathode is at cryogenic temperature in an SRF gun. The effects due to laser heating~\cite{Jared16} can be minor, if the laser fluence can be controlled at a miniature level of (0.1 mJ/cm$^2$). The required driving laser fluence is $F=N_eh\upsilon/\textrm{QE}^{-1}A^{-1}$, where $N_e$ is average number of photoelectrons per pulse, $h\upsilon$ is the drive photon energy, the quantum efficiency $\textrm{QE}=N_e/N_{ph}$ is defined as the ratio between the numbers of photoelectrons and incident photons, and $A$ is the emission area. For example, with $N_e=0.5$, $h\upsilon=4.5$ eV, $\textrm{QE}=1\times10^{-5}$, and $A=\pi(100~\textrm{nm})^2$ (rms size 50 nm), the drive laser fluence is $F=0.12$ mJ/cm$^2$. The laser intensity is $I=F/\tau_\textrm{laser}=6.3$ GW/cm$^2$, where $\tau_\textrm{laser}=18.2$ fs is the laser pulse duration. As the emission area $A$ gets further reduced, a fews effects should be considered, including whether the laser heating is strong enough to increase the intrinsic emittance, whether the absorbed $I$ is less than a few tens of GW/cm$^2$ to avoid multiphoton emission and hence excessive energy spread of photoelectrons, and if the absorbed $F$ is less than a few tens of mJ/cm$^2$ to avoid optical damage.   

\begin{figure}[htb]
\includegraphics[width=0.45\textwidth]{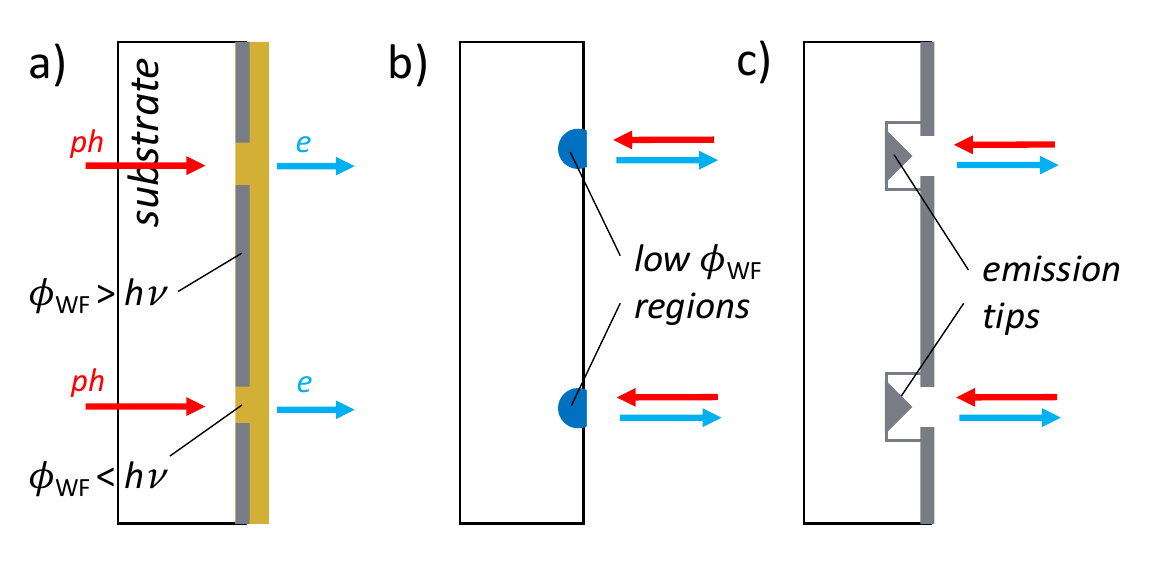}
\caption{(Color online) Illustrations of the cathode concepts for the 'reference beam technique'. Two beams each with pm-rad emittance can be generated with transverse and temporal separations from the photocathode. (a) Back-illuminated flat cathode with aperture masks to define the emission area. (b) Front-illuminated cathode with nano-fabricated nm size, low work function emission sites. (c) Laser trigged photoemission nanotips. The red and blue arrows indicate the trajectory of the cathode drive laser and photoelectrons, respectively. There is time delay between the two laser pulses.}
\label{fig:cathodesolution}
\end{figure}

There are a few promising paths to reduce the emission area to $\sim$100 nm or less to generate pm-rad emittance from a photocathode. The schematics of these concepts are illustrated in Fig.~\ref{fig:cathodesolution}. First, it is feasible to directly focus the cathode driving laser to a spot size similar to its wavelength. With this approach the final focusing optics needs to be very close (within a few mm) to the cathode surface, hence a back-illuminated~\cite{nagoya08,Inagaki14,Lee16,millerfibergun16} and also high gradient rf field compatible photocathode should be used. One step further, one can nano-engineer an aperture on the back side of the cathode to more precisely control the emission area, as shown in Fig.~\ref{fig:cathodesolution}(a). Second, on a flat and uniform cathode surface, assisted by electron or ion beam lithography one can dope a small area to reduce the photoemission work function. Then by tuning the laser wavelength photoelectrons will only be generated from the doped area, as shown in Fig.~\ref{fig:cathodesolution}(b). Third, one can engineer nano-structures to confine optical intensities to sub-wavelength sites through surface plasmons effects, and photoemission will only happen at these high optical intensities regions~\cite{lrkspr13, alexspr13}. While the nano-structures may increase the geometric curvature of the emission surface and induce transverse rf electric fields, which both increase the intrinsic emittance. Finally, multiphoton emission or field-assisted single-photon emission from nanotips also provide nm source size and pm-rad emittance~\cite{hommelhoff15,Feist16,ralph16}. However, above threshold ionization should be avoided, which may otherwise broadens the initial energy spread to several eV. Considering the large local field enhancement close to the apex of the nanotips, the compatibility and robustness of these tips with several tens of MV/m global gradient needs to be experimentally explored and verified. 

\section{effects of electron-electron interactions}
\label{sec6}

In this section, we will discuss how $e$-$e$ interactions affect the energy resolution of the EELS measurement. The interactions between the probe and reference beams can potentially shift the average energy of each individual beam, i.e. introducing uncertainty in the {\it difference} between their energies $\delta|E_\textrm{probe}-E_\textrm{ref}|$. The interactions within each beam, if there contains more than one electron, will significantly broaden the energy spread of the probe beam $\delta E_\textrm{probe}$ and hence degrade the energy resolution. 

\begin{figure}[htb]
\includegraphics[width=0.45\textwidth]{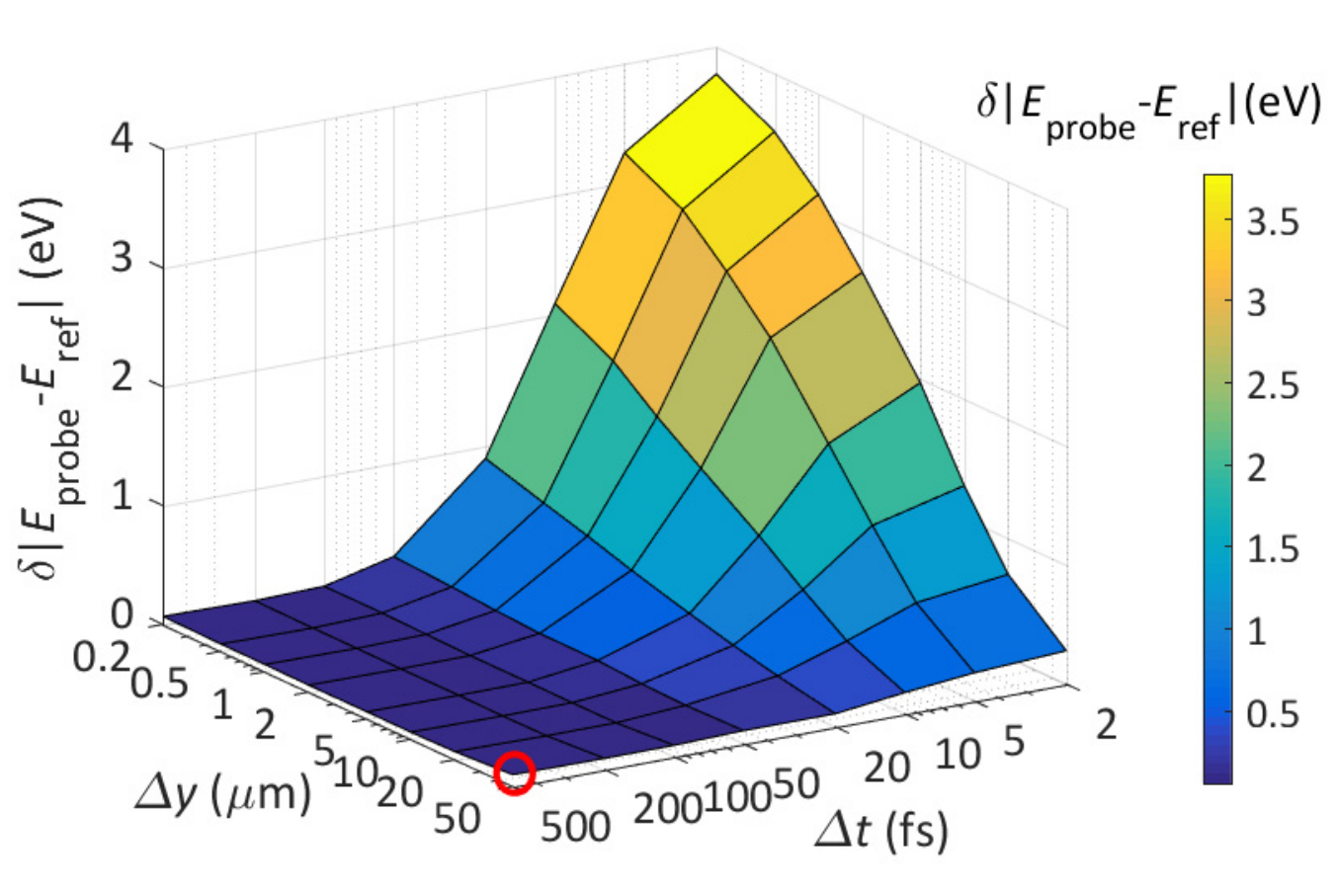}
\caption{(Color online) Uncertainty of $E_\textrm{probe}-E_\textrm{ref}$ for different offsets $\Delta y$ and temporal offsets $\Delta t$ of the probe and reference beams, due to $e$-$e$ interactions with exactly one $e$ in the probe beam and one $e$ in the reference beam. Other beam parameters are summarized in Table.~\ref{tab1}. For each data point, 1000 simulation runs are performed to build up the statistics.}
\label{fig:ediff1e1e}
\end{figure}

We first consider the simplest case: when there is exactly one $e$ in the probe beam and one $e$ in the reference beam. It is straightforward to track the interaction between the two particles from the cathode to the detector. The 'spacecharge3D' algorithm in GPT, which directly calculates relativistic point-to-point interactions, was used. For each simulation run, we randomly generate one $e$ within the defined phase space for the probe beam, and similarly one $e$ for the reference beam. The simulation is repeated multiple times to establish the statistics. The dependence of $\delta|E_\textrm{probe}-E_\textrm{ref}|$ on the initial transverse and temporal offsets $\Delta y$ and $\Delta t$ is shown in Fig.~\ref{fig:ediff1e1e}. $\delta|E_\textrm{probe}-E_\textrm{ref}|$ quickly decreases with larger $\Delta y$ and $\Delta t$. We choose $\Delta y=50~\mu$m and $\Delta t=0.5$ ps where the interaction between the single-$e$ probe beam and single-$e$ reference beam becomes negligible. Note that the result is not divergent as $\Delta y$ and $\Delta t$ are approaching zero, since both of the initial transverse and temporal beam sizes are finite rather than a point. 

The probability for that there are $n$ electrons in the beam follows the Poisson distribution $P(n)=\lambda^ne^{-\lambda}/n!$, where $\lambda$ is the average number of electrons per pulse. It is obvious that even we choose $\lambda$ equal to or less than 1, there is non-zero probability that the beam contains more than one $e$. When there are two $e$ in either of the probe or reference beam, the main effect is a significant growth of the energy spread, while the average energies of the two beams stay approximately constant with offsets $\Delta y=50~\mu$m and $\Delta t=0.5$ ps. The energy spread of a two-$e$ beam can be calculated also in a straightforward way. In each simulation run, two particles are launched randomly within the initial phase space volume of a single beam and tracked from the cathode to the detector. The result can also be extrapolated from Fig.~\ref{fig:ediff1e1e} by pushing both $\Delta y$ and $\Delta t$ to zero. The FW50 energy spread of a two-$e$ beam is 3.3 eV with beam parameters listed in Table.~\ref{tab1}. 

\begin{figure}[htb]
\includegraphics[width=0.45\textwidth]{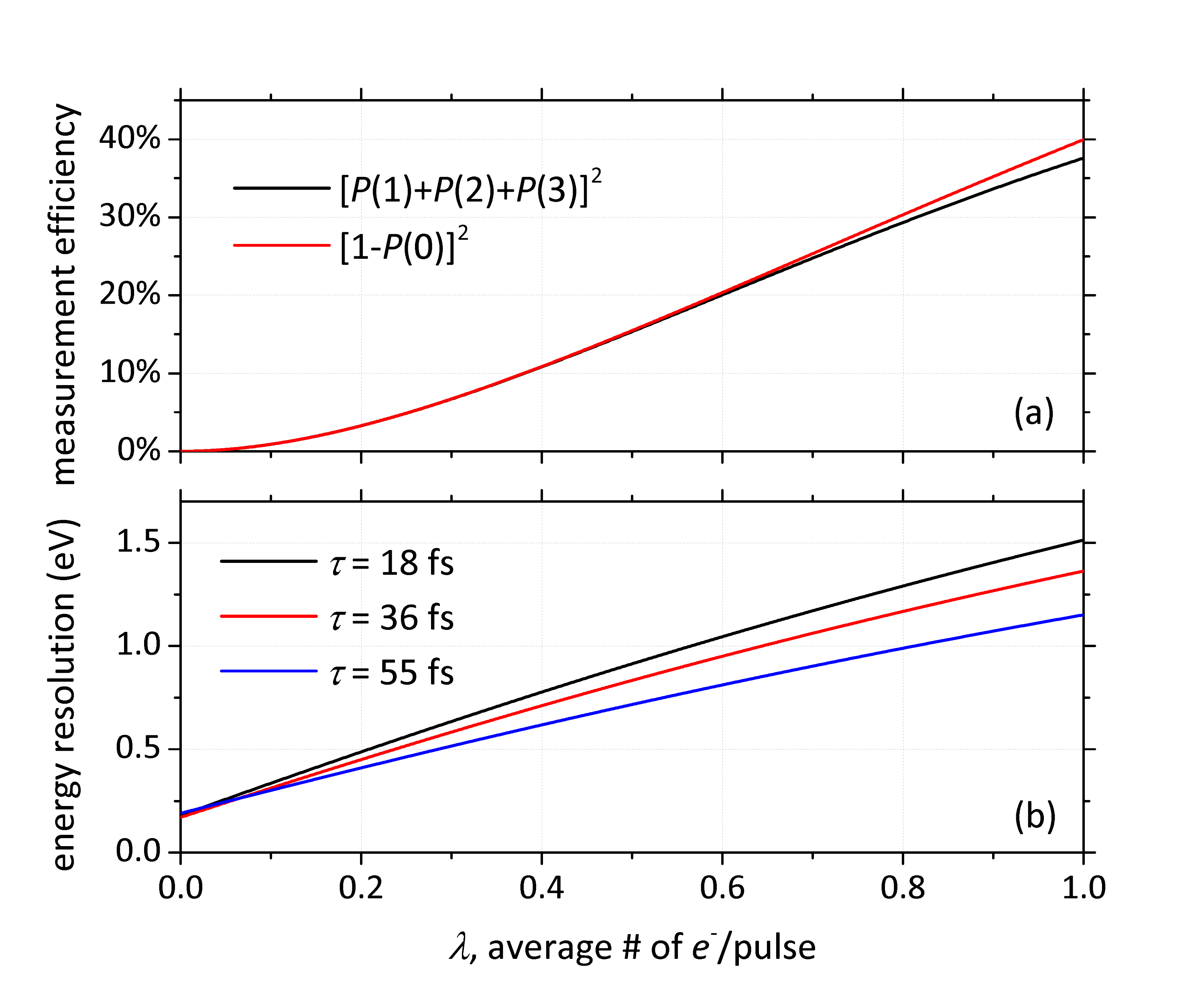}
\caption{(Color online) (a) Measurement efficiency of the 'reference beam technique' due to the Poisson distribution of the number of electrons in the probe and reference beams. The exact efficiency $[1-P(0)]^2$ can be well approximated by $[P(1)+P(2)+P(3)]^2$ for $\lambda<1$. (b) Energy resolution $\delta$ as a function of $\lambda$. One can improve $\delta$ by relaxing the initial bunch length of the electron beam.}
\label{fig:energyres}
\end{figure}

We define the measurement efficiency as the percentage that the probe and reference beams both contain at least one $e$. The measurement efficiency is thus $[1-P(0)]^2$, and can be well approximated by the first few dominating terms $[P(1)+P(2)+P(3)]^2$, as shown in Fig.~\ref{fig:energyres}(a). 

The overall energy resolution of the EELS measurement is the weighted average over all possible combinations of beam charge, e.g. $1e$-$1e$, $1e$-$2e$, $2e$-$1e$, $2e$-$2e$..., for the probe and reference beams. As we discussed before, for the reference beam only its average is relevant to the energy resolution, even if its energy spread grows when containing two or more $e$. Thus the energy resolution will be dominated by the energy spread of the probe beam. We summarize in Table.~\ref{tab4} the energy resolution $\delta_1$, $\delta_2$ and $\delta_3$ when there are 1, 2, and 3 electrons in the probe beam, respectively. The overall energy resolution can be calculated as
\begin{equation}
\delta=\frac{\sum_n\delta_nP(n)}{\sum_n P(n)},
\end{equation}
and the result is shown in Fig.~\ref{fig:energyres}(b). In the limit of $\lambda\rightarrow0$, $\delta$ is $<$0.2 eV and has no contribution from $e$-$e$ interactions. As $\lambda$ increases, the contributions from $P(2)$ and $P(3)$ become more significant and $\delta$ increases. One may notice that if the final overall energy resolution target is notably larger than 0.1 eV, hence no need to maintain 0.1 eV measurement resolution in the spectrometer, is it worth to increases the initial spot size to reduce the effects of $e-e$ interactions? This approach turns out to be not very effective. The reason is that the photoelectrons have 0.2 mrad divergence from the cathode, thus the transverse beam size will soon (within a few 100s of $\mu$m from the cathode) be dominated by the divergence rather than the initial size. Instead, slightly increasing the initial pulse duration is more effective to reduce the $e$-$e$ interactions induced growth of $\delta$. The reason is that the relative longitudinal particle motion is quickly frozen and longer initial pulse duration directly translates into larger spacing between particles. In Table.~\ref{tab4} and Fig.~\ref{fig:energyres} we also show the results with $\times2$ and $\times3$ times longer drive laser pulse duration, which improve the energy resolution by roughly $25\%$.  

\begin{table}[h]
\caption{\label{tab4} Energy resolution $\delta_1$, $\delta_2$, and $\delta_3$ when there are 1, 2, and 3 $e^-$ in the probe beam, respectively. $\delta_1$, $\delta_2$, and $\delta_3$ can be improved by lengthening the initial bunch length, however at the cost of compromised temporal resolution.}
\begin{ruledtabular}
\begin{tabular}{cccc}
(eV) & $\tau=18 \textrm{fs}$ & $\tau=36 \textrm{fs}$ & $\tau=55 \textrm{fs}$ \\
\hline
$\delta_1$ & 0.18 & 0.17 & 0.19 \\
$\delta_2$ & 3.3 & 3.0 & 2.4 \\
$\delta_3$ & 4.1 & 3.5 & 3.1 \\
\end{tabular}
\end{ruledtabular}
\end{table}

\section{summary and discussion}

In this paper we present the concept and design a fs EELS system based on a high gradient, mulit-MeV energy  photocathode rf gun. The tens of MV/m acceleration gradient and several MeV output energy of the rf gun are critical to the generation of 10 fs bunch length, which enables an one-order-of-magnitude improvement of the temporal resolution beyond existing technologies. However, it's a challenge to reach eV level energy resolution, since the energy stability of the electron beams is at best at $1\times10^{-5}$ or 50 eV out of 5 MeV level with state-of-the-art rf amplitude and phase performances. To tackle the challenge, we propose a 'reference beam technique' which can effectively eliminate the effects of the rf instability. 

With the 'reference beam technique', we generate a pair of electron beams, called the 'probe beam' and 'reference beam' respectively, each time from the cathode with controlled spatial and temporal separations. By properly choosing the beam and gun parameters, the energy {\it difference} between the two beams can be precisely controlled and become essentially immune to the rf jitter. Both beams will be recorded by the spectrometer detector on shot-by-shot basis. Only the probe beam will interact with samples and extract the spectroscopic information of the dynamic process, and the average energy of the reference beam serves as the reference to the position of the zero loss peak. We quantitatively studied the requirements on the beam parameters, first-order spectrometer design, and the contribution to the energy resolution from the rf field and $e$-$e$ interactions. Supported by detailed numerical modeling, we demonstrate the feasibility of achieving sub-eV energy resolution and 10 fs-level temporal resolution.

It is worth pointing out that the required key hardware components to realize fs MeV EELS are all under active R\&D. It is promising that the assumed specifications used in our design can be available in near future. For example, there is tremendous effort to improve the gradient of CW superconducting and normal-conducting guns from current 20 MV/m level to $>$40 MV/m for future XFELs~\cite{fes16}. Ultra-stable rf power source and rf-to-laser synchronization system is being developed for these facilities for better stability and temporal control. The key technologies for the high-speed detectors at these facilities can naturally benefit the development of the MHz readout, single-electron sensitivity, $\mu$m spatial resolution detector for the spectrometer. Moreover, although spectrometer detectors usually require $>$1000 pixels in the dispersion direction, but far fewer number of pixels in the vertical direction, thus the total of number of pixels is much less than two-dimensional imaging detectors and it is less challenging to reach higher readout rate. 

A natural extension to the design presented in this paper is to further reduce the probe size to nanometer or even Angstrom scale, which will enable atomic-level spatially column-by-column mapping of electronic dynamics. With an aberration corrected spectrometer one can tolerate much larger beam divergence, hence much stronger focusing to form sharper probe size. At the same time, one should minimize the photoemission area and intrinsic divergence toward a transversely coherent electron source. 

The requirements on smallest possible pulse duration and energy spread, are pushing the limit of the longitudinal emittance of the electron beam. Considering the uncertainty principle for time and energy $\Delta E\Delta t\geq\hbar/2$, with $\Delta E=0.1$ eV FWHM the lower limit for $\Delta t$ is roughly 0.3 fs FWHM. One can approach this limit starting from better understanding and controlling the photoemission process. For a conserved longitudinal emittance, one should explore rf, THz, and optical based beam manipulation for generating attosecond pulse durations or meV energy spread tailored for various applications.  

\begin{acknowledgments}

The authors are grateful to P. Musumeci for helpful discussions. This work was supported in part by the U.S. Department of Energy Contract No. DE-AC02-76SF00515 and the SLAC UED/UEM Initiative Program Development Fund.

\end{acknowledgments}

\nocite{*}


\providecommand{\noopsort}[1]{}\providecommand{\singleletter}[1]{#1}%

\end{document}